# METHODOLOGICAL SOCIETIES

Ammar LAHLOUHI

Department of computer science, University of Batna, 05 000 Batna, ALGERIA

## ABSTRACT

*The evolution of self-adaptive systems poses the problems of their coherence and the resume of the systems' functioning taking into account the accomplished work. While they are the base of the self-adaptive systems, these two aspects are not considered in the related works. In this paper, we propose a methodological based approach. In such approach, the adaptive system's evolution is thought at its model level where its execution is made on the system by exploiting a methodological process. For its concretization, we use colored Petri nets to describe the agents' individual tasks. To handle the system's functioning resume, we exploit the property of Petri nets on which the control flow depends on last marking only.*

## 1. INTRODUCTION

The simultaneous explosion of information and integration of technology together with the continuous evolution from software intensive systems to systems of systems to ultra-large-scale systems requires new and innovative approaches for building, running and managing software systems [12]. The systems are increasingly expected to function in changing environments and to react to changes within the system. As a consequence, the systems are expected to become more versatile, flexible, resilient, dependable, robust, continuously available, energy-efficient, recoverable, customizable, self-healing, configurable, or self-optimizing by adapting to changing requirements and contexts/environments [10]. One of the most promising approaches to achieve such properties is to equip systems with self-adaptation capabilities.

Self-adaptive systems are of primary importance, especially, for the environments that are:

1. very unstable and/or unforeseeable, such as military combat terrain,
2. unknown and/or inaccessible to humans, such as space and spitfire,
3. etc.

However, the development of self-adaptive systems is far from reaching necessary maturity for a massive engineering. Self-adaptive systems have to make decisions on adaptability at runtime with respect to changing requirements [13]. Self-adaptation means the ability of the system to modify its behavior and/or structure in response to its perception of the environment and the system itself, and to its requirements [11]. Self-adaptive system is capable of evaluating and changing its own behavior, whenever the evaluation shows that it is not accomplishing what it was intended to do, or when better functionality or performance may be possible [14]. Two problems are then particularly embarrassing for handling the evolution of adaptive systems:

1. The possibility that the adaptation leads to an incoherent system,
2. The system functioning resume after its evolution can induce to losing what is already accomplished.

The works on self-adaptive systems considers the coherence of the resulted system in an empirical way while they discard the functioning resume. In this paper, we adopted a methodology based approach. Although the general approach can be generalized to all systems' models, we apply to multi-agent systems. Our objective, in this paper, is to focus on some necessary characteristics of adaptive systems' evolution.

The remainder of the paper is organized in 3 sections. In section 2, we describe the two main evolution's difficulties of the systems adaptation, which are a coherent evolution and system's functioning resume. In the section 3, we explain methodological approach to the systems adaptation in multi-agent systems context. Section 4 presents an application of the approach to a methodological society, a system that oscillates between development and deployment. Lastly, we finish this paper by a conclusion in which we show some advantages and also some disadvantages of this work.

## 2. METHODOLOGY BASED SELF-ADAPTIVE SYSTEMS

A system is always a realization of some model. Such model will be implicit in the empirical development whereas it is explicit in the methodical development. We build a system by using a process of models' implementation. The model itself is built using a design process making it possible to develop it starting from an objective laid down for the system (see figure 1). We qualify the process of design and implementation as a methodological process.

The system's adaptation is done as follows:

1. Stop the functioning of the old system S to be adapted,
2. Modify, by a coherent evolution, S to produce a new system S',
3. Restart the functioning of S' by recovering the work achieved by S. The recovery of accomplished work is qualified of functioning resume.

This section is organized in three subsections. In subsection 2.1, we explain the coherence of the system's evolution and how we take it into account by a methodology based approach. In the subsection 2.2, we explain the functioning resume of an adaptive system and the approach based on the recovery of the system's control flow and its state used to overcome this problem. We finish this section by the characterization of the self-adaptation and the requirements of an evolution of self-adaptive systems.

### 2.1 Coherence of the System's Evolution

The coherence of the system's evolution is the conformity of its functioning to the objective which is laid down to it, i.e., its functioning must make it possible to achieve such objective. In the works on adaptive systems, the coherence of the system's evolution was not tackled in a methodical way. This has some disadvantages such as:

1. The systems' adaptation cannot be justified in a clear manner. What doesn't simplify its discussion in a transparent way,
2. A non-methodical approach cannot ensure that the intermediate versions of the adaptive system are coherent.

In this paper, we bind the systems' adaptation to the methodological process of its development. This makes it possible to produce systems which are coherent compared to their objectives provided that the methodological process used is correct and can be implemented by the system. Moreover, the evolution's operations can be justified and discussed.

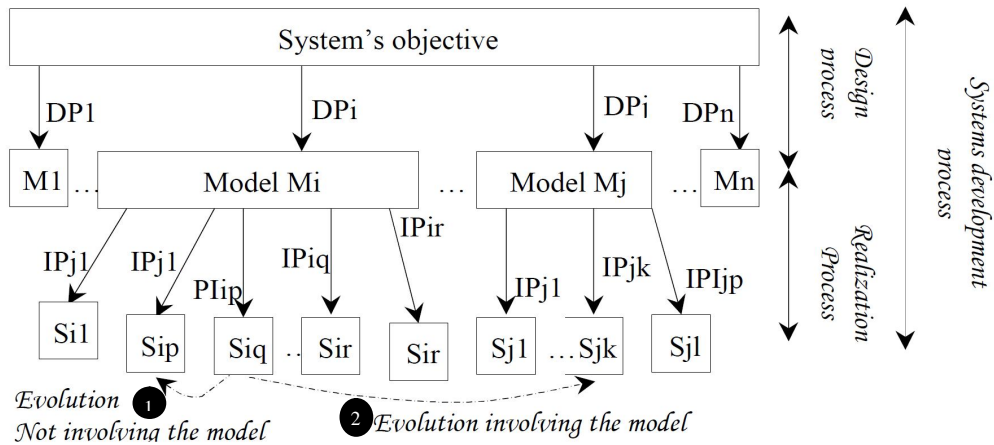

**Fig. 1.** The relation of the system's evolution to its development process (DP: Design Process, IP: Implementation Process, M: Model, S: System)

As shown in figure 1, after an adaptation of a system $S_{iq}$ (of a model $M_i$), it becomes another system. The latter can be the result of one of the following two realizations:

1.  Technical adaptation: It doesn't involve the system's model.
2.  Model based adaptation: It involves the model. It is a realization $S_{jk}$ of another model $M_j$.

The technical evolution can be seen as another realization Sip of the same model $M_i$ without modifying the relevant aspects described in the model. This is possible since the model is an abstraction where the system's evolution can include aspects not considered in its model. It is a different implementation of the same model using the same implementation process that is used to build $S_{iq}$ or another implementation process.

The model based adaptation is a design of another model $M_k$ followed by an implementation of $M_k$ into $S_{kl}$ (see Figure 1). It can also be regarded as an evolution of the model $M_i$ of the system $S_{ij}$ towards another model $M_k$ followed by the implementation of $M_k$ into a $S_{kl}$ (see mE association in Figure 2). In figure 2, it comes to reverse $S_{ij}$ to its model $M_j$ and then evolve $M_j$ to a new model as shown in Eq. (1).

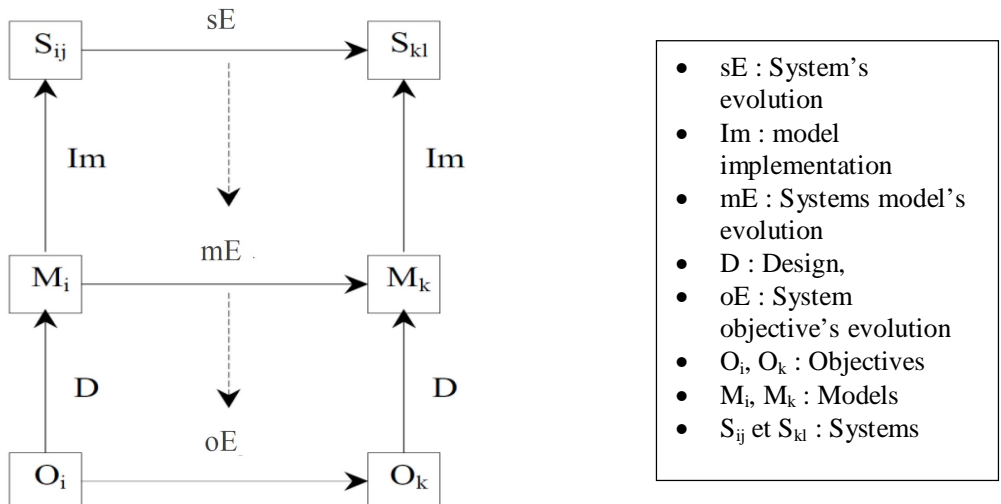

**Fig. 2.** Relation between the system's evolution, and its models and objectives.

We consider a first class, which is that of model-aware systems. A system of such class (maS, see Eq. (2)) holds its model where $Im^{-1}(S_{ij}) = M_i$. For such systems, we can use Eq. (3) instead of Eq. (1).

Now, we consider a second class that is of self-evolving systems aware of their models. Such systems (seS, see Eq. (4)) should include operations (sE, specified in equation (3)) of their coherent evolution, which are of two kinds:

1. Operations of a coherent evolution of the system's model to another model (mE),
2. An implementation process of models (Im).

$$sE(S_{ij}) = Im(mE(Im^{-1}(S_{ij}))) \qquad (1)$$

$$maS = (S, M) \qquad (2)$$

$$sE(S_{ij}) = Im(mE(M_i)) \qquad (3)$$

$$seS = (maS, mE, Im) \qquad (4)$$

The implementation process doesn't always include a complete reimplementation of the model but rather it will include a process identifying the elements that will be modified and the sub-process to be implemented to modify consequently the system. However, the evolution operations mE must be taken from the implementation process of the used methodology. In section 3, we give an example of such modifiable elements with the associated sub-processes.

## 2.2 Functioning Resume

The system's adaptation during its operation must take into account the work accomplished before its evolution. For that, the new system's functioning must resume the functioning of the old system. In spite of its importance, this problem was not tackled in the works on adaptive systems. The approach proposed in this paper is based on the following process:

1. The system S stops the work in progress,
2. Save the context C of the system S (Its state and its control flow) before S's adaptation,
3. Adapt S into S', as described in 2.1,
4. Start again S' while considering the context C.

This process raises two difficulties:

1. S' cannot support the context C,
2. The contexts' management is not obvious.

For the support of the context, we distinguish between the plan and the system that will implement it. Such a distinction makes it possible to subdivide the adaptation problem into two kinds of adaptations. If the adaptation modifies the plan, the S' cannot resume the previous work and, consequently, it initiates another one. If the adaptation doesn't modify the plan, it is then necessary to consider what was accomplished of the plan and to continue the remaining part of such plan.

The contexts' management is not evident. Moreover, it depends on the physical systems where their operating systems can be heterogeneous. Such heterogeneity leads to heterogeneous approaches and upped implementation information at the design level. The approach adopted in this paper is based on the colored Petri nets marking. We will present such approach in subsection 3.2.2.

Finally, to be able to resume its functioning, a self-adaptive system must be aware of its context (Ct). It will then have knowledge about such context (C), and it will be able to save C (sC), and to resume (rC) its functioning using C, as described in Eq. (5).

$$Ct = (C, sC, rC) \qquad (5)$$

## 2.3 Methodological Self-adaptive Systems

As given in Eq. (6), a self-adaptive system must be self-evolving and aware of its context. Its elements are summarized in Eq. (7). It will have knowledge about:

1. Its model M,
2. The evolution's operations of the systems' models in a coherent way and, in particular, those of the evolution of its model (mE),
3. The implementation process (Im) of its model,
4. Its context (C) and how it saves it (sC) and how it resumes (rC) such context,
5. How to take again its functioning so that it takes into account C.

$$saS = (seS, Ct) \qquad (6)$$

$$saS = (S, M, mE, Im, C, sC, rC) \qquad (7)$$

A self-adaptive system doesn't stop before adapting but rather it changes its functioning from its functioning to an adaptation state and vice-versa. The algorithm of such system is summarized in figure 3. In section 3, we give details on how we apply such ideas on multi-agent systems modelling; particularly, we give details on functions sC, rC, mE and Im.

```
Algorithm adaptiveSystem(S, M, mE, Im, C, sC, rC){
   While not systemEnded(C) {
      repeat
         neededAdaptation = Go(S, C); // Go a step
         over the system S functioning
      Until (neededAdaptation.sate OR
      systemEndded(C));
      If neededAdaptation.state {
         C=sC();
         S = Im(mE(M, neededAdaptation));
         rC(C);
      }
   }
}
```

Figure 3. Outline of adaptive systems' algorithm

## 3. METHODOLOGY BASED SELF-ADAPTIVE MULTI-AGENT SYSTEMS

In this section, we show a concretization of the methodology based approach in the multi-agent context. In subsection 3.1, we describe a particular multi-agent model and development methodology and then, in subsection 3.2, we clarify how we concretized the methodology based approach using such model and development process.

## 3.1 Multi-agent Systems Development

### 3.1.1 Organization

An organization's structure is composed of elements and a relation of communication between them. In reality, we build an organization to achieve a given objective. From the methodological viewpoint, we put, in our mind a goal and we build then the organization that will make it possible to reach such a goal. Obviously, the goal will be merged in the organization once it is built. However, methodologically, it is present. We define then an organization as a pair (Organization's structure, Global objective).

We can replace the declarative expression "objective" by the procedural expression "reach the objective". This later is the organization's task. We redefine, consequently, the organization as a pair (organization's structure, global task).

In this paper, we employ colored Petri nets CPN (see [3, 4], for details on theoretical foundations and their practical use of CPN) for describing the global task. We made such description as follows:

1. A transition contains a role attached to one of its knowledge's procedure,
2. Any entry arc to a transition is labeled by a procedure's input,
3. Any exit arc from a transition is labeled by a procedure's output,
4. Places are communication means between these procedures (outputs of some procedures and inputs of others).

### 3.1.2 Agents and Multi-agent Systems

A multi-agent system is a whole composed of agents, which are systems assuming roles in a given organization. The agents are cooperatives in the sense that their objectives are sub-objectives of an organization's global objective. The agents' objectives are then derived from those of the organization according to a roles' attribution. Since the organizations' global objectives are described using CPN, the objectives of the agents will be also expressed using CPN.

In Petri nets, the communication is implicit. This makes difficult to express the communication in the individual tasks of the agents' models. Research works, on the modeling of the agents' behaviors, use autonomous Petri nets where they present the communications by arcs (see [9], for example). This does not express clearly the individual tasks' description; we then employ synchronized (non-autonomous) CPN (see [8], non-autonomous Petri nets). In the autonomous Petri nets, the transitions are subject to the internal conditions only, in particular, places' marking. In addition to these internal conditions, the transitions in the non-autonomous Petri nets are subject to external events also. In this paper, we regard the communication's messages as events attached to the transitions. An outgoing arc of a transition schematizes the message's sending while an entering arc schematizes its reception.

### 3.1.3 A Multi-Agent Methodological Process

In an organization development of multi-agent systems, the methodological process consists in:

1. Build an organization that allows satisfying the system's requirements. This includes:

    a. a description of the organization's global task,
    b. the organization's communication structure,

2. Derive a multi-agent model that must satisfy the organization. This includes:

    a. roles attribution to the agents,
    b. derivation, according to this attribution, of the individual tasks and other elements (including sensors and effectors for communication purpose) of the agents from the organization's global task.

3. Implementation of software agents' models as a distributed object oriented system.

We consider the previous methodological process as the implementation process (Im function of Eq. (7)). Details on its concretization in real practical case study are given in section 4.

### 3.1.4 Conclusion

By following the multi-agent development methodology described in 3.1.3, we can derive multi-agent models and then agents' models. An agent model is composed of:

1. Sensors and effectors including those allowing the agent to interact with its environment but also those allowing it to communicate with other agents,
2. Knowledge that is a set of needed procedures so that the agent can accomplish its individual task,
3. The agent's individual task.

The sensors and effectors are implemented as objects and the knowledge procedures are implemented as methods of objects classes (in the sense of object-oriented models). The agent's execution is an execution of its individual task. This execution is made by invoking its knowledge procedures.

Finally, note that, to be able to initialize the multi-agent system's functioning, we need a starting system to launch the various software agents.

## 3.2 Use of the Methodology Based Approach

In this subsection, we show how one can identify the evolution's operations of the developed multi-agent model. We detail also the approach used for the functioning resume by exploiting the CPNs' properties.

In the subsection 2.3, we identified the requirements of a system to be self-adaptive. In addition to be aware of its model and the implementation process, an adaptive multi-agent system must have, as showed in Eq. (7):

• knowledge enabling it to evolve multi-agent models in a coherent way (mE),
• the ability to save and restore its context (sC and rC),
• an ability for the functioning resume.

In subsection 3.2.1, we identify the evolution's operations. In subsection 3.2.2, we explain the approach of saving and restitution of the context for the functioning resume.

### 3.2.1 Coherent Evolution of Multi-agent Models

A multi-agent model consists in a set of agents' models and a communication relation between them. Consequently, the evolution's operations (mE) of such a model concern:

1. Add, remove and modify a model of the set of agents' models,
2. Add (aCom), remove (rCom) and replace (rpCom) a communication of the communication relations between the agents,

An agent model modification can involve:

1. Add (aS), remove (rS) and replace (rpS) a sensor,
2. Add (aE), remove (rE) and replace (rpE) an effector,
3. Add (aP), remove (rP) and replace (rpP) a knowledge procedure,
4. replacement of the individual task (rpT).

Finally, we resume the operations of multi-agent model's evolution in Eq. (8).

$$mE = (aCom, rCom, rpCom, aSn, rSn, rpSn, aEf, rEf, rpEf, aP, rp, rpP, rpT) \quad (8)$$

### 3.2.2 Context Saving and Functioning Resume

The element of the agent's model concerned with the functioning resume is its behavior. The modeling of the agent's behavior is made using CPNs. Consequently, the agent's functioning resume is that of the continuity of the CPN's interpretation. We made this by:

1. Saving the CPN's marking before the system's adaptation (sCPN),
2. Adapt the multi-agent system, as explained in subsection 3.2.1,
3. Restore the CPN's marking by considering it as an initial marking (rCPN),
4. Continue the interpretation of the agent's behavior.

The CPN's marking (mCPN) will, simply, be loaded and its interpretation launched whereas the agent resume its execution. The association between the system's context (Eq. (5)) is given in Eq. (9). This simple approach benefices from the CPN's characteristics of portability (independence from the computers' technologies and programming languages) and robustness (benefits from the CPN's theoretical foundation).

$$Ct = (C, sC, rC) = (mCPN, sCPN, rCPN) \quad (9)$$

### 3.2.3 Conclusion

Finally, we lead to the Eq. (10) which summarizes the composition of the adaptive multi-agent system according to the methodological based approach.

$$saS = \begin{pmatrix} S, M, aCom, rCom, rpCom, \\ aSn, rSn, rpSn, \\ aEf, rEf, rpEf, \\ aP, rp, rpP, \\ rpT, \\ Im, \\ mCPN, sCPN, rCPN \end{pmatrix} \quad (10)$$

## 4. METHODOLOGICAL SOCIETY

In this section, we show how we applied the methodological approach in an example of what we called methodological society (a multi-agent system for a methodology). This example is a part of software development processes and, in particular, their application to software engineering environment. The software development processes knew a significant turn in the middle of the eighty years with Osterweil's paper [7] entitled "Software processes are software too" which became the slogan of the software process's community. Osterweil considered the software process as being a complex software system where its development must follow the software development process itself. Since this date, several projects were led (MARVEL [5], EPS [2], ADEL [1], etc.).

In our example, we consider the software development process as a multi-agent system. We then use our approach for its development. We developed then a cooperative and distributed environment, as a multi-agent system functioning. The process was described entirety as a CPN of an organization's global task.

This environment includes:

1. An automation of sub-processes, such as the derivation of the necessary knowledge of the agents from the global task's description, etc.,
2. A colored Petri nets editor including a module of global tasks decomposition into agents' individual tasks according to a roles attribution,
3. A module of multi-agent systems synthesis starting from the multi-agent model.

This is a first step towards an adaptive environment for multi-agent systems development. However, several tasks of software development cannot be completely automated. The humans intervention is then of primary importance for the achievement of some tasks, such as formulation of the global task and the roles attribution. This intervention must be well planned and well controlled. For that, we used the cooperation based approach [6] for humans' integration in the multi-agent systems. The software agents will have then in charge programmed routines, embarrassing tasks, whereas we reserve for human agents the tasks requiring intelligence and/or actions on the environment external to the multi-agent system.

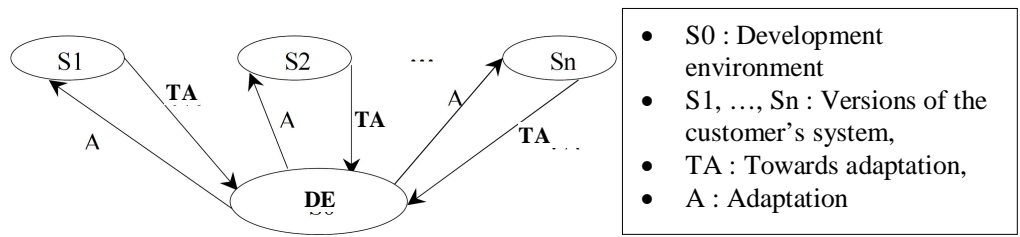

**Fig. 5.** Adaptive system's evolution

An adaptive environment, developed according to our approach, can be seen from various viewpoints. It is the case, for example, of an environment which oscillates from methodological process system to become the software system to be developed and vice-versa. In this case, once the environment system receives the first requirements it works to produce a new multi-agent system (MAS). It adapts then to become the new MAS to satisfy the objective of such MAS by introducing, if needed, new agents and/or withdrawing others. At the need, the new MAS self-adapt in turn (see figure 5) to become a development environment to modify the MAS which will adapt again to become another new MAS.

Such vision makes it possible to organize the agents to belong to various societies on a world wide area network including those of the development environment. The systems development society develops systems' models and its agents leave the development environment by entering the necessary agents of the developed MAS. Once the needs for the system changes, there would be an intervention of the agents of the development society and the system's agents leave the MAS.

## 4. CONCLUSION

The work described in this paper is currently at hand. The proposed approach is based not only on the agents' adaptability in an individual way but on the adaptability of the multi-agent system while not excluding the agents' adaptability which is seen like a particular case. What is not the case of several works in this field which consider the adaptability of the multi-agent system as that of the individual agents only by endowing such agents with the mechanisms allowing them to self-adapt. We can see the adaptation in such works as agent's knowledge adaptation. Another advantage of our approach, in addition to the motivation of this work discussed along the paper, is that of the openness of the multi-agent system where agents can enter and leave coherently the system.

The approach has however some disadvantages such as the need for an automation of some parts of the methodological process. What is, generally, not always possible. This requires the human intervention in the adaptation process. However, these aspects are those which raise more difficulties in other works and which remained open where such difficulties find some answers in our approach. Once the system development will reach some maturity, some of these problems can find automatic solutions. Lastly, it is this human intervention which widened the approach's applicability. The adaptive systems not raising major difficulties can be founded on automatable parts of the methodological process whereas the others must also include manual parts.

Since the agents are autonomous, the adaptation of a multi-agent systems can be done by parts, i.e., while some agents of the multi-agent system are self-adapting, other agents continue to run normally. We plan to explore such alternative in the near future.

## REFERENCES

[1]   Belkhatir, N., Estublier, J., & Melo, W. L. (199). Activity coordination in Adele: a software production kernel. Proc of the 7th International software process Workshop, IEEE Computer Society Press, San Francisco, CA, October 16-18, 1991. (pp. 48-50).

[2]   Ciancarini, P. (1993). Coordinating rule-based software processes with ESP. ACM Transactions on Software Engineering and Methodology (TOSEM), 2(3), 203-227.

[3]   Jensen, K. (1994). An introduction to the theoretical aspects of coloured petri nets (pp. 230-272). Springer Berlin Heidelberg.

[4]   Jensen, K. (1998). An introduction to the practical use of coloured petri nets. In Lectures on Petri Nets II: Applications (pp. 237-292). Springer Berlin Heidelberg.

[5]   Kaiser, G. E., Barghouti, N. S., & Sokolsky, M. H. (1990). Preliminary experience with process modeling in the Marvel software development environment kernel. In System Sciences, 1990., Proceedings of the Twenty-Third Annual Hawaii International Conference on (Vol. 2, pp. 131-140). IEEE.

[6]   Lahlouhi, A., Sahnoun, Z., Benbrahim, M. L., & Boussaha, A. (2002). Interface agents development in MASA for human integration in multiagent systems. In Advances in Artificial Intelligence—IBERAMIA 2002 (pp. 566-574). Springer Berlin Heidelberg.

[7]   Osterweil, L. (1987). Software processes are software too. In Proceedings of the 9th international conference on Software Engineering (pp. 2-13). IEEE Computer Society Press.

[8]   David, R., & Alla, H. (2010). Non-Autonomous Petri Nets. In Discrete, Continuous, and Hybrid Petri Nets (pp. 61-116). Springer Berlin Heidelberg.

[9]   Xu, H., & Shatz, S. M. (2001). A framework for modeling agent-oriented software. In Distributed Computing Systems, 2001. 21st International Conference on. (pp. 57-64). IEEE.

[10]  Cheng, B. H.C. et al. (2009). Software engineering for self-adaptive systems: A research roadmap. In Software engineering for self-adaptive systems (pp. 1-26). Springer Berlin Heidelberg.

[11]  De Lemos, R., et al. (2013). Software engineering for self-adaptive systems: A second research roadmap. In Software Engineering for Self-Adaptive Systems II (pp. 1-32). Springer Berlin Heidelberg.

[12] Gabriel, R. P., Northrop, L., Schmidt, D. C., & Sullivan, K. (2006). Ultra-large-scale systems. In Companion to the 21st ACM SIGPLAN symposium on Object-oriented programming systems, languages, and applications (pp. 632-634). ACM.

[13] Weiss, G., Becker, K., Kamphausen, B., Radermacher, A., & Gérard, S. (2011). Modeldriven development of self-describing components for self-adaptive distributed embedded systems. In (pp. 477–484).

[14] Salehie, M., & Tahvildari, L. (2005). Autonomic computing: emerging trends and open problems. In Proceedings of the 2005 workshop on Design and evolution of autonomic application software (pp. 1–7). St. Louis, Missouri: ACM.